\documentclass[twocolumn,prx,aps,superscriptaddress,longbibliography]{revtex4-2}

\usepackage[utf8]{inputenc}
\usepackage{amsmath,amssymb}
\usepackage{graphicx}
\usepackage{floatrow}
\usepackage[T1]{fontenc}
\usepackage{physics}
\usepackage{tikz,bm}
\usepackage[caption=false]{subfig}
\usepackage{epigraph}
\usepackage{appendix}
\usepackage{hyperref,cleveref}

\newcommand{\be}{\begin{equation}} 
\newcommand{\ee}{\end{equation}}

\begin{document}
\title{Non-equilibrium quantum probing through linear response}
\author{S. Blair}\email{sblair13@qub.ac.uk}
\affiliation{Centre for Quantum Materials and Technologies, School of Mathematics and Physics, Queen's University Belfast, University Road, Belfast BT7 1NN, United Kingdom}
\author{G. Zicari}\email{gzicari01@qub.ac.uk}
\affiliation{Centre for Quantum Materials and Technologies, School of Mathematics and Physics, Queen's University Belfast, University Road, Belfast BT7 1NN, United Kingdom}
\author{A. Belenchia}
\affiliation{Institut f\"ur Theoretische Physik, Eberhard-Karls-Universit\"at T\"ubingen, 72076 T\"ubingen, Germany}
\affiliation{Centre for Quantum Materials and Technologies, School of Mathematics and Physics, Queen's University Belfast, University Road, Belfast BT7 1NN, United Kingdom}
\author{A. Ferraro}
\affiliation{Centre for Quantum Materials and Technologies, School of Mathematics and Physics, Queen's University Belfast, University Road, Belfast BT7 1NN, United Kingdom}
\affiliation{Dipartimento di Fisica Aldo Pontremoli, Universit\`{a} degli Studi di Milano, I-20133 Milano, Italy}
\author{M. Paternostro}
\affiliation{Centre for Quantum Materials and Technologies, School of Mathematics and Physics, Queen's University Belfast, University Road, Belfast BT7 1NN, United Kingdom}
\affiliation{Università degli Studi di Palermo, Dipartimento di Fisica e Chimica—Emilio Segrè, via Archirafi 36, Palermo I-90123, Italy}
\date{\today}

\begin{abstract}
The formalism of linear response theory can be extended to encompass physical situations where an open quantum system evolves toward a non-equilibrium steady-state. Here, we use the framework put forward by Konopik and Lutz [Phys. Rev. Research {\bf 1}, 033156 (2019)] to go beyond unitary perturbations of the dynamics. Considering an open system comprised of two coupled quantum harmonic oscillators, we study the system's response to unitary perturbations, affecting the Hamiltonian dynamics, as well as non-unitary perturbations, affecting the properties of the environment, e.g., its temperature and squeezing. We show that linear response, combined with a quantum probing approach, can effectively provide valuable quantitative information about the perturbation and characteristics of the environment, even in cases of non-unitary dynamics.

\end{abstract}

\maketitle
 
\section{Introduction}

The study of non-equilibrium scenarios is key for the dynamics -- and thermodynamics -- of quantum systems~\cite{deVega:2017,ThermoBook:2018}. Although an overarching framework to consistently interpret the variety of non-equilibrium quantum phenomena is hitherto missing, some light has been shed towards understanding several aspects of this multifaceted problem. These theoretical studies come together with a strong interest in applications concerned with transport phenomena, e.g., in quantum technologies at the nanoscale~\cite{Benenti:2017}.

Given a non-equilibrium setting, the standard approach follows classical statistical mechanics~\cite{Marconi:2008}, where one applies small perturbations to the system of interest to gather information about the system itself: depending on the nature of the perturbation, one would be able to deduce relevant physical properties through, e.g., response and relaxation functions or generalized susceptibilities~\cite{kubobook}.
This simple, yet insightful, idea is the essence of linear response theory; this being usually the first step in trying to harness the otherwise complex phenomenology of non-equilibrium systems. However, it is worth emphasising that, when extended to quantum systems, Kubo's original formulation of linear response theory relies on two assumptions~\cite{Kubofdt,Kubolinresp}: the system is isolated so that its dynamics are unitary; furthermore, it relaxes towards a thermal equilibrium state. These conditions are not always met in practice. On one hand, closed quantum systems are usually a crude approximation, as we should effectively include environmental effects in the form of dissipation and/or decoherence~\cite{breuer,Rivas2012}. On the other hand, there are cases in which the system relaxes towards a non-equilibrium steady-state (NESS). This class of states is obtained, for instance, when considering \emph{boundary-driven} systems, i.e., systems that are dissipatively driven by coupling with two different baths at the two ends~\cite{LandiNonEqReview}. The physical properties of the system and its bath (e.g., temperature or chemical potential), as well the intra-system and system-bath coupling will ultimately influence the transport properties. 

In this work, we aim to probe the non-equilibrium dynamics of a system through linear response theory. To this end, we extend the customary domain of application of linear response theory to include non-unitary perturbations to the system dynamics. Non-unitary perturbations have been considered in Ref.~\cite{MehboudiAcin}, where a method was put forward for finding the linear response of the covariance matrix to a perturbation in the Gaussian channel describing its evolution. The perturbation could be either unitary or non-unitary. Here we proceed via a different method where we find the linear response of an observable of the system to a perturbation in the master equation describing its evolution. As a paradigmatic setting of boundary-driven systems, we consider a simple yet highly non-trivial example and address two coupled harmonic oscillators locally interacting with their own bath. The non-vanishing interaction between the two parties eventually leads the composite system to a NESS. Furthermore, we assume that the dynamics of such a system are described by a local master equation which does not include memory effects, i.e., we rely on a fully Markovian description of the evolution~\cite{breuer}.
This scenario has been analysed in Ref.~\cite{lutz}, where, inspired by considerations coming from classical statistical mechanics~\cite{Agarwal:1972,Seifert:2010,Baiesi:2013}, a framework for non-equilibrium quantum linear response has been put forth. However, consistently with the original spirit of the linear response approach, the interest is usually limited to unitary perturbations ~\cite{lutz, Levi:2021} --- e.g., to the interaction Hamiltonian between the two subsystems~\cite{lutz}. 

Here we use one of the two oscillators to test the effects of the perturbations to the bath that is locally coupled to the other one. This configuration is fully in line with the quantum probing paradigm, where it is customary to use a small, simple, and controllable system interacting with a more complex environment to infer accurate information about environmental parameters (e.g., temperature or spectral properties)~\cite{Smirne:2013,Benedetti:2018,Tamascelli:2020,Strunz:2021}. In turn, the problem that we address is closely connected with quantum estimation theory~\cite{Paris:2009,Giovannetti:2006,Polino:2020,Sidhu:2020,thermometryBelfast}, for which the ultimate goal is to find the optimal measurement scheme to gain precise information about a set of given parameters characterising the system and its dynamics~\cite{Stace:2010,Correa:2015,DePasquale:2018,Mehboudi:2019,Mitchison:2020}. Interestingly, within the context of non-equilibrium open quantum systems described by Markovian dynamical maps, a close relationship between fluctuation-dissipation theorems~\cite{Campisi_rev:2011} and quantum metrology has been established~\cite{Mehboudi2018}, though limited to the static case: such a connection allows for the description of the effect of a perturbation within the linear response regime.

We first investigate the case where the probe itself is in contact with a thermal bath; despite the thermal noise, we are able to detect the changes to the environment interacting with the main system, when a sudden quench is applied to its temperature. In line with the general aim of quantitatively assessing the effects of a sudden change of the macroscopic parameters characterising the dissipation, we also consider the case where the system interacts with a thermal squeezed bath, while we assume that the probe is perfectly isolated.
We show that linear response extended to the case of non-unitary dynamics remains effective in providing key information on the perturbation affecting the system and, when paired with a quantum probing approach, offers a valuable quantitative tool for the characterisation of the features of an environment. 

The remainder of the paper is organized as follows. In Sec.~\ref{sec:linresptheory} we set out the theory of linear response of open quantum systems to small perturbations. We try to keep our formalism as general as possible, so that both unitary and non-unitary perturbations can be applied to the system dynamics.  In Sec.~\ref{sec:SystemThermalBaths}
we describe our first system of interest: a pair of coupled quantum harmonic oscillators, each interacting with local thermal environments. Given the nature of the systems and states that we consider, we can solve the dynamics analytically using the formalism of Gaussian quantum mechanics without the need for numerical approximations. In Sec.~\ref{sec2perturb}
we apply two simultaneous perturbations -- one unitary and one non-unitary. We then focus on the special case of a perturbation of the bath temperature. In \ref{seceffecton2}
we show that the second oscillator can serve as a probe of the response to a perturbation on the first oscillator. In Sec.~\ref{sec:sqbath}, we detach one of the two subsystems (which becomes our probe) from the corresponding local bath, while we let the other subsystem interact with a local squeezed thermal bath. We then study the response of the probe to perturbations on the parameter controlling the squeezing of the bath. In Sec.~\ref{sec:conclusions}, we present our conclusions, along with an outlook for future directions.

\section{Linear response in open quantum systems}
\label{sec:linresptheory}

In this Section, we will review the formalism of linear response for open quantum systems. In particular, we will explicitly show that the linear response of a generic observable to a perturbation can be written in terms of the steady-state of the system and the observable of interest, written in the Heisenberg picture, where the time-evolution is solely controlled by the unperturbed dynamics \cite{lutz}. Since our aim is to investigate how the system responds whenever we perturb either the unitary or the non-unitary part of the dynamics, we will write the relevant equations in a general form. Thus we will work at the level of superoperators, with the Liouvillian of the dynamics expressed as the sum of a unitary part and a dissipator in the standard Lindbladian form \cite{breuer,Lindblad:1976,Gorini:1976,Gorini:1978}.  The dynamics of the system are described by the Markovian master equation
\be
\dot{\rho}(t)={\cal L}\rho(t)\,,
\label{eq:markovME}
\ee
where the Liouvillian $\mathcal{L}$ can be expanded (to first order) as
\be
{\cal L}={\cal L}_0+\epsilon(t){\cal L}_1+{\cal O}(\epsilon^2)\,,
\ee
where ${\cal L}_0\rho(t)=-i[H_0,\rho(t)]+{\cal D}\rho(t)$ with $H_0$ the unperturbed Hamiltonian and ${\cal D}$ the system's dissipator. Also, $\epsilon = \epsilon(t)$ is the time-dependent parameter controlling the perturbation. Following lines similar to those in Ref.~\cite{lutz}, we assume that, for a fixed value of $\epsilon$, the dynamics possess a stationary state $\rho_\epsilon$, i.e., ${\cal L}\rho_\epsilon=0$, that can be expressed as 
\be
\rho_\epsilon=\rho_0+\epsilon \rho_1+{\cal O}(\epsilon^2)\,,
\ee
where $\rho_0$ is the steady-state of the unperturbed dynamics \footnote{In the following we use what in Ref.~\cite{lutz} is defined as a class one quantum response function, involving only the stationary state of the unperturbed dynamics. It is interesting to note that, to this end it is sufficient to assume that there exists a stationary state of the unperturbed dynamics.}, and $\rho_1$ its linear (in $\epsilon$) correction due to the perturbation. It is worth emphasising that $\rho_\epsilon$ is not necessarily an equilibrium state. When that is the case and the perturbation is unitary, we recover the standard Kubo formalism of linear response \cite{Kubolinresp}.

Moreover, in general, the state at some time $t$ can be expressed as
\be
\rho(t)=\rho_0+\pi_1(t)\,,
\label{eq:rho(t)}
\ee
where $\pi_1(t)$ has to be traceless in order for $\rho(t)$ to be a physical state. By substituting Eq.~(\ref{eq:rho(t)}) into Eq.~(\ref{eq:markovME}) and integrating over time we arrive at the final result 
\be
\pi_1(t)=\int_0^\tau dt \,\epsilon(t)e^{{\cal L}_0 (\tau-t)}{\cal L}_1\rho_0\,,
\ee
to first order in $\epsilon$ \cite{lutz, zwanzig2001noneq}. It follows that the linear response of a generic observable $A$ can be written as
\be
\label{eq:linear_response_DeltaA}
\overline{\Delta A(\tau)}=\langle A \rangle _\tau - \langle A \rangle _0 = \int_0^\tau dt \, \epsilon(t) {\cal R}_A (\tau-t)\,,
\ee
where $\langle A \rangle _0$ is the unperturbed expectation value and ${\cal R}_A(t)=\textrm{Tr}[A e^{{\cal L}_0 t}{\cal L}_1\rho_0]$ is the linear response function. Note that this response function can be expressed as 
\be
\label{eq:response_schro}
{\cal R}_A(t)=\textrm{Tr}[e^{{\cal L}^*_0 t} [A] {\cal L}_1\rho_0]\equiv\textrm{Tr}[A_H(t) {\cal L}_1\rho_0]\,,
\ee
where $\mathcal{L}_0^*$ is the dual of the generator ${\cal L}_0$ of the unperturbed dynamics,
and $A_H(t)$ evolves according to the adjoint master equation \footnote{Here and in the remainder of the manuscript, unless otherwise specified, we will assume units such that $\hbar=1$ and the Boltzmann constant $k_B=1$.}
\begin{equation}
    \dot{A}_H(t)=i[H_0,A_H(t)]+{\cal D}^*\left[A_H(t)\right]
\label{eqAdot}
\end{equation}
that encompasses the dual dissipator ${\cal D}^*[\cdot]$~\cite{breuer}.

\section{System in contact with local thermal baths}
\label{sec:SystemThermalBaths}

\begin{figure}
\center{\includegraphics[scale=0.37]{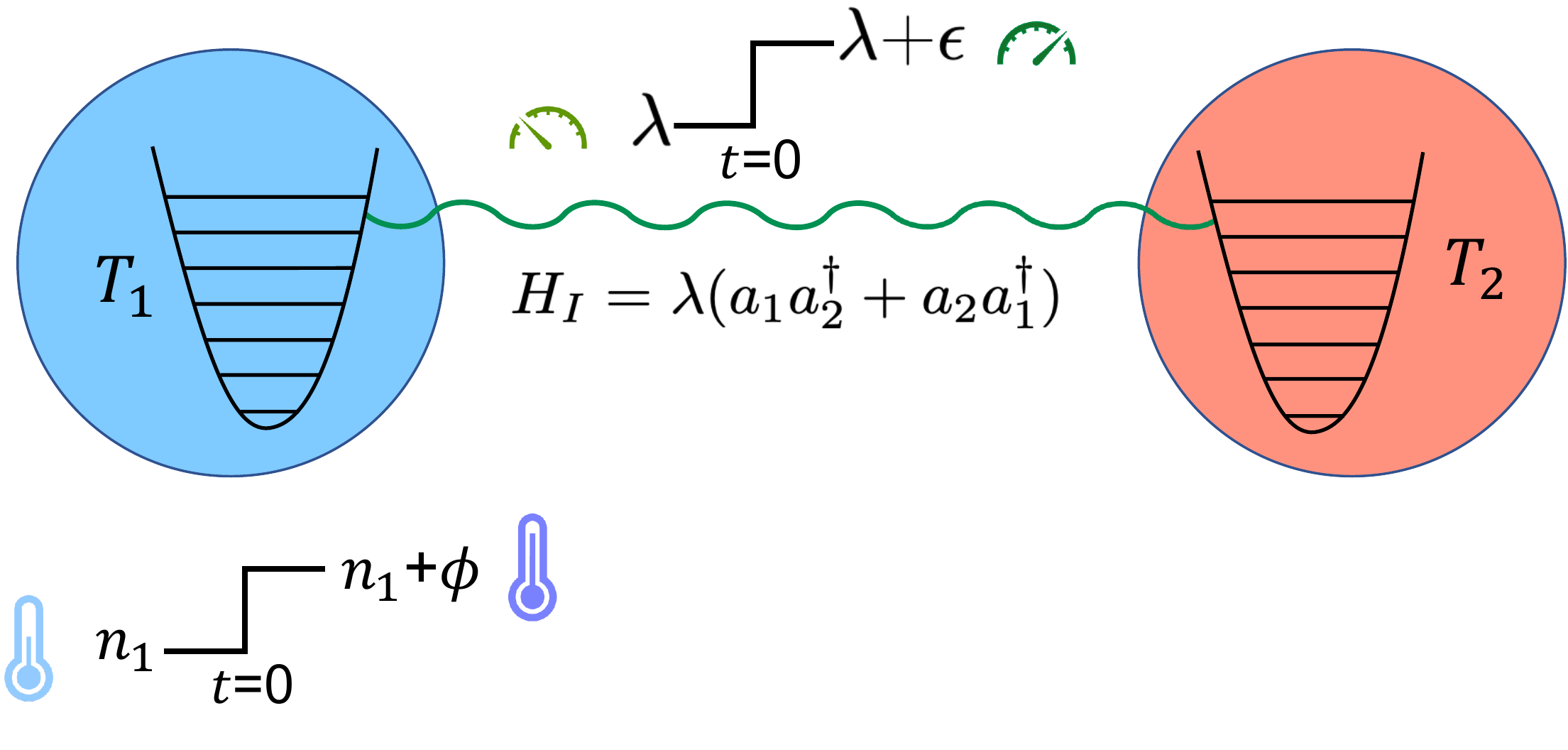}}
    \caption{Sketch of the system investigated in Sec.~\ref{sec:SystemThermalBaths}. It consists of two coupled quantum harmonic oscillators, interacting with their local thermal baths at two different temperatures, $T_1$ and $T_2$. We study the response of one of the two subsystems when we apply a sudden quench either to the coupling constant (i.e., $\lambda \to \lambda + \epsilon$) or to the number of excitations in the first bath (i.e., $n_1 \to n_1 + \phi$).}
    \label{fig:2oscdiagram}
\end{figure} 

First, let us consider a system of two coupled quantum harmonic oscillators, each interacting with a local thermal bath, as depicted in Fig.~\ref{fig:2oscdiagram}. The unperturbed Hamiltonian of this system reads
\begin{equation}
\label{h0}
    H_0 = \omega_1 a_1^\dag a_1+\omega_2 a_2^\dag a_2+\lambda(a_1 a_2^\dag + a_2 a_1^\dag)\,,
\end{equation}
where $\omega_1$ is the frequency of the first harmonic oscillator and $\omega_2$ is that of the second, which is detuned by a quantity $\delta$, i.e., $ \omega_2 \equiv \omega_1+\delta$. The interaction between the two oscillators, controlled by the coupling strength $\lambda$,  is modelled so as to preserve the total number of excitations across the closed system.

The open dynamics of the unperturbed system are governed by the master equation 
\be
\dot{\rho}={\cal L}_0(\rho)=-i[H_0,\rho]+\sum_{j=1,2}{\cal D}_j[\rho] \,,
\label{eq:unpertme1}
\ee
where each local dissipator reads ($j=1,2$)
\begin{equation}
\begin{aligned}
\label{eqmaster}
    {\cal D}_j[\rho]&=\gamma\,(n_j +1)\left( a_j \rho a_j^\dag -\frac{1}{2}\{a_j^\dag a_j, \rho\}\right)\\
    & + \gamma\, n_j\left( a_j^\dag \rho a_j -\frac{1}{2}\{a_j a_j^\dag, \rho\}\right).
\end{aligned}
\end{equation}
Each $\mathcal{D}_j[\rho]$, defined in terms of the local creation and annihilation operators $a_j^\dagger$ and $a_j$, is the sum of two terms: the first describes the incoherent loss of excitations, while the second describes incoherent pumping. Here, $\gamma$ is the damping rate (assumed to be the same for both oscillators), while the mean number of excitations for the $j$-th thermal bath at inverse temperature $\beta_j=1/{T_j}$ is $n_j=\left[\exp(\beta_j\omega_j)-1\right]^{-1}$. 

Given that the Liouvillian of the unperturbed dynamics is quadratic in the creation and annihilation operators, we can employ the methods of Gaussian quantum mechanics to solve the dynamics exactly. This is a valid approach as far as we restrict our considerations to Gaussian states and, as we will do in the following, to perturbations that are also quadratic forms of the operators of the oscillators. Hence we only need to study the evolution of the first and second moments of the dimensionless position and momentum operators $x_i=(a_i+a_i^\dag)/\sqrt{2}$ and $p_i=(a_i-a_i^\dag)/(i\sqrt{2})$. Given the vector 
of quadratures $\mathbf{Y}\equiv (x_1,p_1,x_2,p_2)$, 
 and starting from Eq.~(\ref{eq:unpertme1}), it is straightforward to check that the first moments $\bar{Y}_i \equiv \langle Y_i \rangle$ are damped to zero~\cite{gaussianstates}. Therefore, if we assume -- without loss of generality -- the first moments to be initially zero, the system dynamics will be solely determined by the second moments. This is a customary assumption, e.g., when one considers optomechanical set-ups, where the Gaussian modes always represent zero-mean fluctuations around some semi-classical steady-state~\cite{VitaliMirrorField,reconstructing}. 

The second moments are encoded in the covariance matrix (CM) $\boldsymbol{\sigma}$, whose entries are defined as $\sigma_{ij}=\frac{1}{2}\langle\{Y_i,Y_j\}\rangle-\langle Y_i \rangle \langle Y_j \rangle$.
The time-evolution is thus expressed in the Lyapunov form as $\dot{\boldsymbol{\sigma}}={\bm\alpha} \boldsymbol{\sigma}+\boldsymbol{\sigma} {\bm\alpha}^T +\mathbf{D}$,
where ${\bm\alpha}$ and $\mathbf{D}$ are the drift and diffusion matrices, respectively. While the former bears dependence upon both the unitary and the non-unitary (dissipative) part of the dynamics, the latter only depends on the non-unitary part, describing the effect of thermal noise due to the interaction between the system and the environment.
In order to obtain the linear response of an observable $A$ of the system as given in Eq.~(\ref{eq:linear_response_DeltaA}), we need the steady-state of Eq.~(\ref{eq:unpertme1}) and the Heisenberg-evolved observable $A_H(t)$. The steady-state $\rho_0$ is completely characterized by the solution to ${\bm\alpha} \boldsymbol{\sigma}+\boldsymbol{\sigma} {\bm\alpha}^T =-\mathbf{D}$, which can be found to read as [cf. Appendix~\ref{app:ss1} for the details of the calculations] 
\begin{equation}
\label{eqcovmatrix_compact}
\boldsymbol{\sigma}_0 =
\zeta
\begin{pmatrix}
\boldsymbol{\Xi}_1 & \boldsymbol{\Theta} \\
\boldsymbol{\Theta}^{\rm T} & \boldsymbol{\Xi}_2
\end{pmatrix},
\end{equation}
where $ \zeta=(\gamma^2+\delta^2)/(4\lambda^2+\gamma^2+\delta^2)$, $\boldsymbol{\Xi}_i = \operatorname{diag} \left ( B + n_i + 1/2, B + n_i + 1/2 \right )$ and
\be
\boldsymbol{\Theta}= C
\begin{pmatrix}
- \delta  & - \gamma  \\
\gamma & - \delta
\end{pmatrix}
\ee
with $B=2\lambda^2(n_1+n_2+1)/(\gamma^2+\delta^2)$, and $C=\lambda(n_1-n_2)/(\gamma^2+\delta^2)$. While, in general, the coupling between the two oscillators results in a non-thermal steady-state, in the limit $\lambda \to 0$, we have $B\to0$, $C\to0$, and $\zeta \to 1$, thus recovering the case of two independent quantum harmonic oscillators, each relaxing towards the corresponding thermal state.

We start by considering the observable $A_1=\beta_1\omega_1 a_1^\dag a_1$. The corresponding adjoint master equation [cf. Eq.~(\ref{eqAdot})] leads to a set of coupled differential equations, whose solution gives us
\begin{align}
\label{eqa1daga1_maintxt}
( a_1^\dag a_1) (t)&  =  f(t)a_1^\dag a_1 + j(t)a_2^\dag a_2 \nonumber \\
& + p(t)a_1a_2^\dag +q(t)a_1^\dag a_2 +s(t).
\end{align}
The explicit forms of the functions $f(t), j(t), p(t), q(t)$, and $s(t)$ are given in Appendix~\ref{app:TimeEvolution}.

\subsection{Two perturbations: coupling strength and temperature}
\label{sec2perturb}
Given the unperturbed dynamics in Eq.~(\ref{eq:unpertme1}), we investigate the response to two perturbations, $\mathcal{L}_1^\epsilon$ and $\mathcal{L}_1^\phi$: the former  is a sudden quench in the coupling strength  $\lambda$ between the two harmonic oscillators, while the latter is a quench in the number of excitations $n_1$ of the first bath. Note that the linear response to a Hamiltonian perturbation of the coupling strength has been thoroughly studied in Ref.~\cite{lutz}. The perturbed master equation reads
\be
\dot{\rho}={\cal L}_0\rho+\epsilon (t){\cal L}_1^\epsilon\rho+\phi (t){\cal L}_1^\phi\rho\,.
\ee
We take 
$\mathcal{L}_1^\epsilon\rho=-i[H_P,\rho]$, where $H_P$ is a beamsplitter-like Hamiltonian perturbation of the form \cite{Aspelmeyer}
\begin{equation}
\epsilon (t) H_P=\epsilon \, \theta(t) (a_1 a_2^\dag + a_2 a_1^\dag)   \, ,
\label{hp}
\end{equation}
where $\epsilon > 0$, and $\theta(t)$ is the Heaviside step function.  
 
We obtain the linear response of observable $A_1$ due to to the Hamiltonian perturbation $\mathcal{L}_1^\epsilon$ as
\be
\overline{\Delta A_1^\epsilon(\tau)}=i\,\epsilon\int_0^\tau dt  \, \text{Tr}\Big[[H_P,A_1(\tau-t)]\rho_0\Big] \, .
\ee
Since $A_1(\tau-t)= \beta_1 \omega_1( a_1^\dag a_1) (\tau - t)$, we can use Eq.~(\ref{eqa1daga1_maintxt}), and rewrite the expectation values in terms of entries of the steady-state CM from Eq.~(\ref{eqcovmatrix_compact}). We get 
\begin{equation}
     \overline{\Delta A_1^\epsilon(\tau)} = \frac{\,2\,\Delta n\,\lambda\,\beta_1\omega_1\epsilon}{z^2(\gamma^2+z^2)} \int_{0}^{\tau}dt\,{\cal G}(\tau-t)e^{-\gamma(\tau-t)},
\label{eqcoupresp}
\end{equation}
where we have introduced the kernel function 
\begin{equation}
    {\cal G}(t)=\gamma\left[4\lambda^2\cos(zt)+\delta^2\right]+(\gamma^2+\delta^2)z\sin(zt)
\end{equation}
with $\Delta n=n_2-n_1$ and $z^2=\delta^2+4\lambda^2$.

Consider now a quench in the temperature of the environment of the first oscillator through a perturbation in $n_1$ described by
\begin{equation}
\begin{aligned}
    \phi (t)\mathcal{L}_1^\phi \rho &= \gamma \: \phi (t)\left[ a_1 \rho a_1^\dag -\frac{1}{2} \Big\{a_1^\dag a_1,\rho\Big\}\right. \\
  &  +\left.a_1^\dag \rho a_1 -\frac{1}{2} \Big\{a_1 a_1^\dag,\rho\Big\}\right],
\label{eqL1}
\end{aligned}
\end{equation}
where $\phi (t)= \phi\, \theta(t)$, with $\phi>0$. The corresponding linear response is
\begin{equation}
    \overline{\Delta A_1^\phi(\tau)} = \int_{0}^{\tau} dt\,\phi(t)\, \text{Tr}\left[\:A_1(\tau -t)\mathcal{L}_1^\phi \rho_0\:\right].
\label{eqexpecs1}
\end{equation}
After some algebra, we find 
\begin{equation}
      \overline{\Delta A_1^\phi(\tau)} = \; \beta_1 \omega_1 \gamma\phi \int_{0}^{\tau} dt \: f(\tau-t)\, ,
\label{eqtempresponse}
\end{equation}
with $f(t)={e^{-\gamma t}}[\delta^2+2\lambda^2+2\lambda^2\cos(zt)]/z^2$ [cf. Appendix~\ref{app:TimeEvolution}]. Finally, the linear response to a combination of perturbations to both the coupling strength of the interaction and the temperature of the first oscillator's bath is obtained by simply adding the two aforementioned contributions.
 Therefore, the linear response (of the energy of the first oscillator) to both perturbations is
\be
   \overline{\Delta A_1(\tau)} = \sum_{k=\epsilon,\phi}\overline{\Delta A_1^k(\tau)}\,,
\label{eq:twopertresp}
\ee
where $\overline{\Delta A_1^\epsilon(\tau)}$ and $\overline{\Delta A_1^\phi(\tau)}$ are given by Eqs. (\ref{eqcoupresp}) and (\ref{eqtempresponse}), respectively.

\begin{figure}[t]
\center{\includegraphics[width=\textwidth]{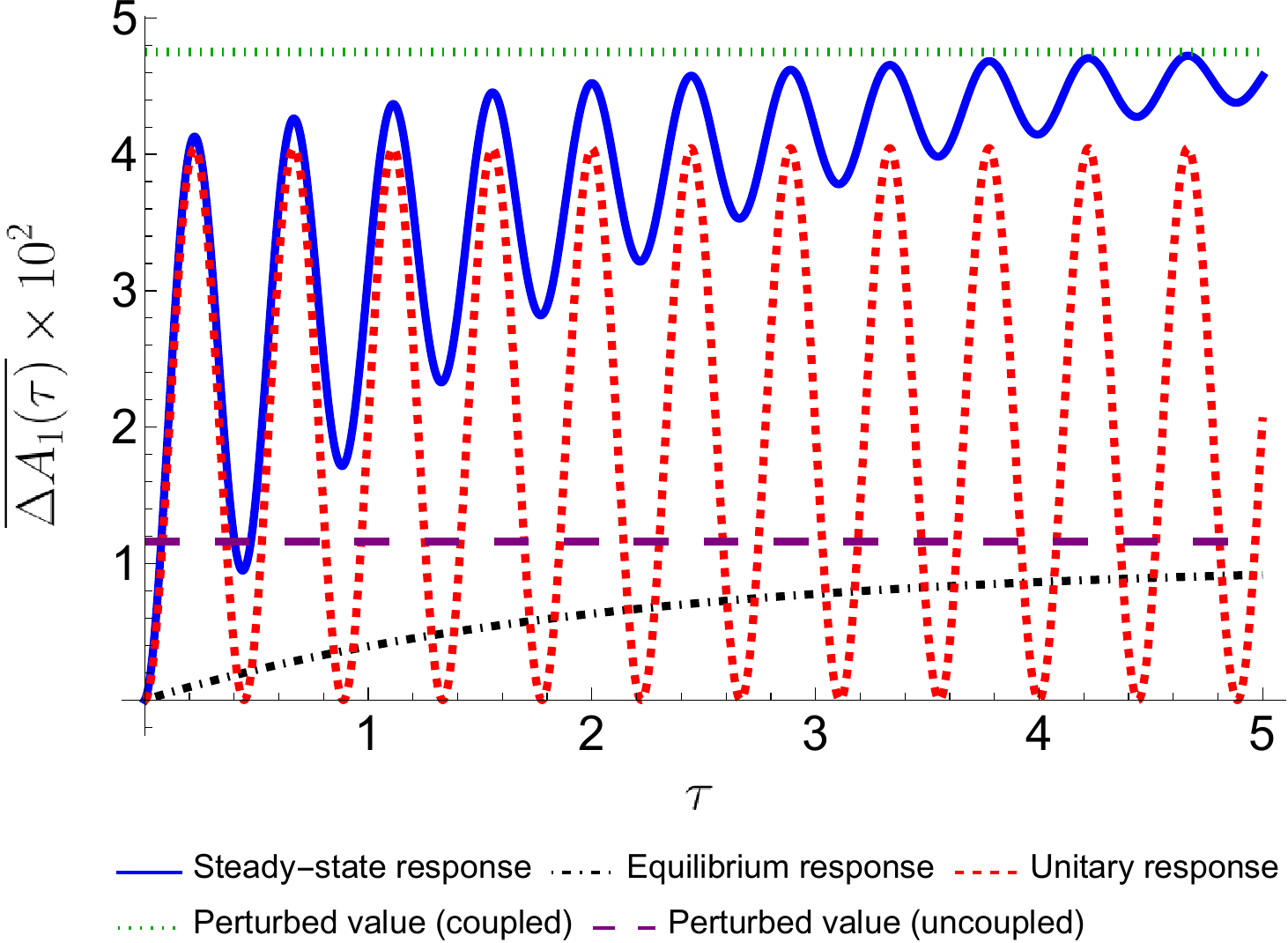}}
\caption{Plot of the dimensionless energy response of the first oscillator (with $A_1=\beta_1\omega_1a_1^\dag a_1$) to a step perturbation to both the coupling strength $\lambda$, and the number $n_1$ of excitations of the first bath. The dynamics are simulated for the following values of the parameters: $\delta=10\omega_1$, $\lambda=5\omega_1$, $\gamma=0.5\omega_1$, $\epsilon=0.1\omega_1$, $\phi=0.1$, $\beta_1\omega_1=0.1$, and $\beta_2\omega_1=0.001$. The steady-state response converges to the perturbed value. The equilibrium response shows what happens in the limit of decoupled oscillators ($\lambda \to 0$), while the unitary response ($\gamma \to 0$) reflects the case where the system is closed.}
\label{2perturbplot}
\end{figure}

In Fig.~\ref{2perturbplot}, we show this response as a function of time  for a certain choice of the parameters of the system (see caption) and with perturbation strengths $\epsilon=0.1\omega_1$ and $\phi=0.1$.  We can also compare the linear response result with the exact dynamics of the perturbed system. In order to do so, we consider the steady-state $\rho'_0$ reached asymptotically by the system when setting the coupling strength to $\lambda+\epsilon$ and the average number of excitations in the bath of the first oscillator to $n_1+\phi$, i.e., taking the perturbations into account. The difference between this new steady-state and the original unperturbed steady-state shall be referred to as the perturbed value of the energy response of the oscillator. We first calculate the expectation value of $A_1$ over the steady-state $\rho_0'$, i.e., $\langle A_1 \rangle^\infty_{\rho_0'}=\lim_{t\to\infty}\beta_1\omega_1\text{Tr}[a_1^\dag a_1(t)\rho_0']$,
as we are looking for asymptotic conditions. The invariance of the steady-state under the dynamics allows us to remove the time dependence, thus leaving
\begin{equation}
    \langle A_1 \rangle_{\rho_0'}^\infty= \beta_1\omega_1\text{Tr}[a_1^\dag a_1\rho_0']=\frac{\beta_1\omega_1}{2}\left[(\sigma'_{11})_0+(\sigma'_{22})_0-1\right]\, ,
\label{eq:pertval2}
\end{equation}
where $(\sigma'_{ij})_0$ are the elements of the CM associated with $\rho_0'$. From this, we need to subtract the expectation value calculated over the unperturbed steady-state, for which $\epsilon=\phi=0$, i.e., $\langle A_1 \rangle^\infty_{\rho_0}=\frac{\beta_1\omega_1}{2}\left[(\sigma_{11})_0+(\sigma_{22})_0-1\right]$. Using the CM given in Eq.~(\ref{eqcovmatrix_compact}) we can find the explicit form of the perturbed value of the energy response of the first oscillator, and compare it with the long time limit of Eq.~(\ref{eq:twopertresp}). It is straightforward to show that the two expressions coincide to first order in $\epsilon$ and $\phi$, consistently with the linear response theory approximation. In Fig.~\ref{2perturbplot}, we clearly see that the perturbations cause oscillations in the steady-state response. The amplitude of such oscillations is damped over time, until the linear response converges to the perturbed value as $t \to \infty$.

We also consider the case of $\lambda \to 0$ where the oscillators are uncoupled and only interact with their own local bath. As 
each oscillator is in its own equilibrium state just before they are perturbed, we refer to this as the equilibrium response. Additionally, the case in which the system evolves unitarily can be recovered from Eq.~(\ref{eq:twopertresp}) by taking $\gamma \to 0$. 
As expected, in this limit, we obtain reversible dynamics, so that the linear response displays recurrences over time, which essentially follow the energy flowing back and forth from one oscillator to the other.

\begin{figure}[t]
\center{\includegraphics[width=\textwidth]{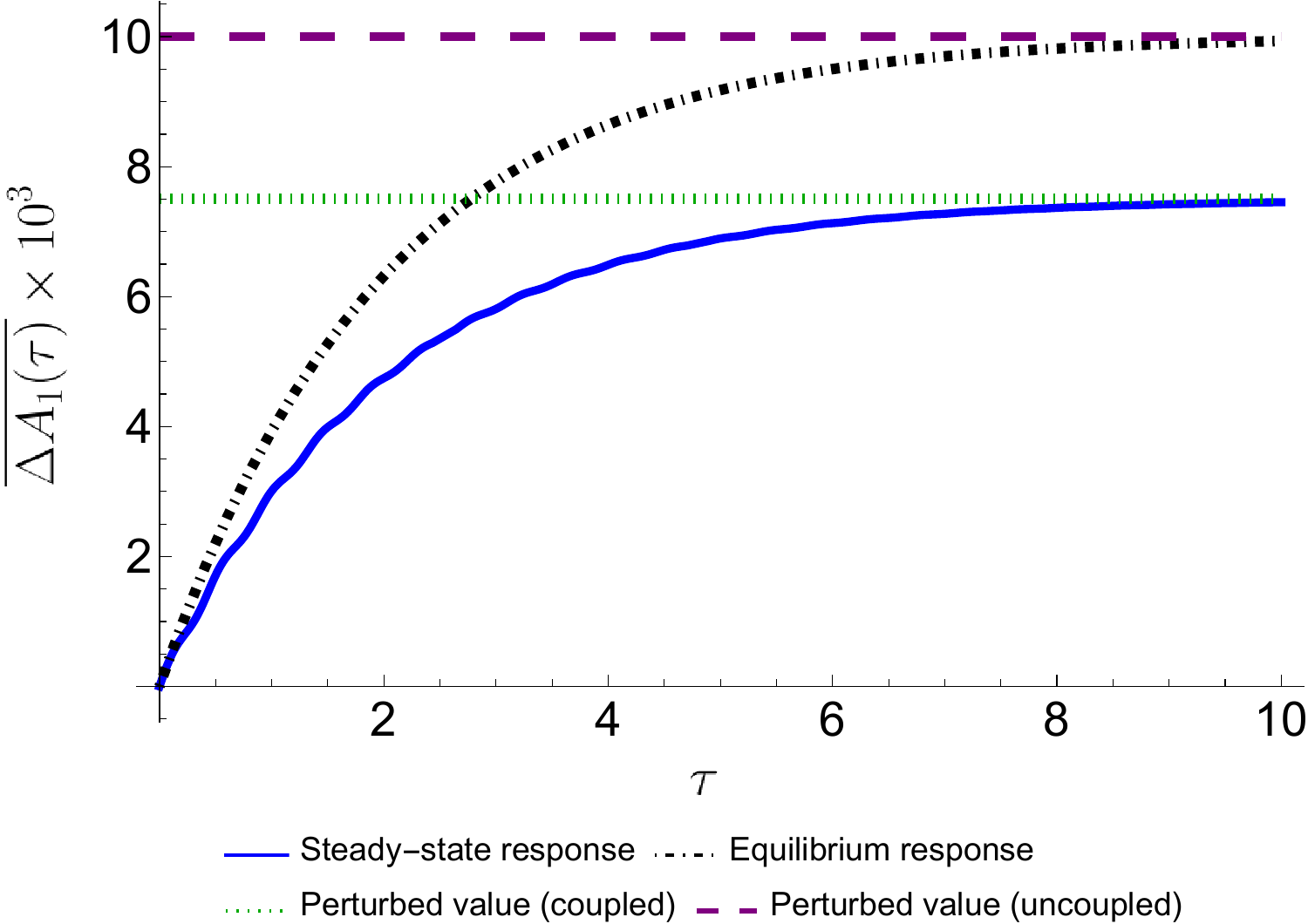}}
\caption{Plot of the linear response of the dimensionless energy of the first oscillator (with $A_1=\beta_1\omega_1a_1^\dag a_1$), for the case of a step perturbation to the number of excitations $n_1$ of the bath locally interacting with it. The plots are obtained with the following choice of the various parameters: $\delta=10\omega_1$, $\lambda=5\omega_1$, $\gamma=0.5\omega_1$, $\phi=0.1$, $\beta_1\omega_1=0.1$, and $\beta_2\omega_1=0.001$. The steady-state response approaches the coupled perturbed value ($\lambda \neq 0$), whilst the equilibrium response ($\lambda \to 0$) approaches its own perturbed value for the case where the oscillators are not coupled.}
\label{tempplot}
\end{figure}

We now analyse a special case of the previous scenario, where we perturb only the temperature of the first bath, while we keep the coupling strength constant. Hence, the response is given by Eq.~(\ref{eqtempresponse}), or, equivalently, by setting $\epsilon=0$ in Eq.~(\ref{eq:twopertresp}). The results are plotted in Fig.~\ref{tempplot}. In this case, where the perturbation affects only temperature, the linear response fully captures the exact dynamic response, which is calculated in Appendix.~\ref{app:exactresp}. The equilibrium response here is identical to the equilibrium response in the case of two perturbations, consistently with the fact Eq.~(\ref{eqcoupresp}) vanishes for $\lambda\to0$, reducing the case of two perturbations to just the response to the quench in temperature. In the equilibrium scenario, we have $f(t) = e^{-\gamma t}$, thus Eq.~(\ref{eqtempresponse}) yields $\overline{\Delta A_1^0(\tau)} = \beta_1\omega_1\phi (1 - e^{- \gamma \tau})$, which, as $\tau \to \infty$, converges to the asymptotic value $\overline{\Delta A_1^0(\infty)} = \beta_1\omega_1\phi$. This is the maximum response possible, and it occurs when $\lambda \to 0$, i.e., when the oscillators are uncoupled. By contrast, whenever we switch on the coupling between the two oscillators, i.e., $\lambda \ne 0$, the response of the first oscillator is comparatively smaller, as -- through the interaction with the second oscillator -- the first oscillator additionally experiences the effects of the second bath.

\subsection{ Second oscillator as a probe of a perturbation}
\label{seceffecton2}
Let us now consider the effect of a perturbation to the temperature of the first bath on the energy of the second oscillator. In order to do so, we apply a quench to the number of excitations in the first bath, and investigate the possibility of gaining information about this perturbation through the second oscillator, thus probing the environment of one subsystem through the response of the other. The linear response is once more given by Eq.~(\ref{eqexpecs1}) with  $A_2=\beta_2\omega_2a_2^\dag a_2$ replacing $A_1$.

\begin{figure}[t]
\center{\includegraphics[width=\textwidth]{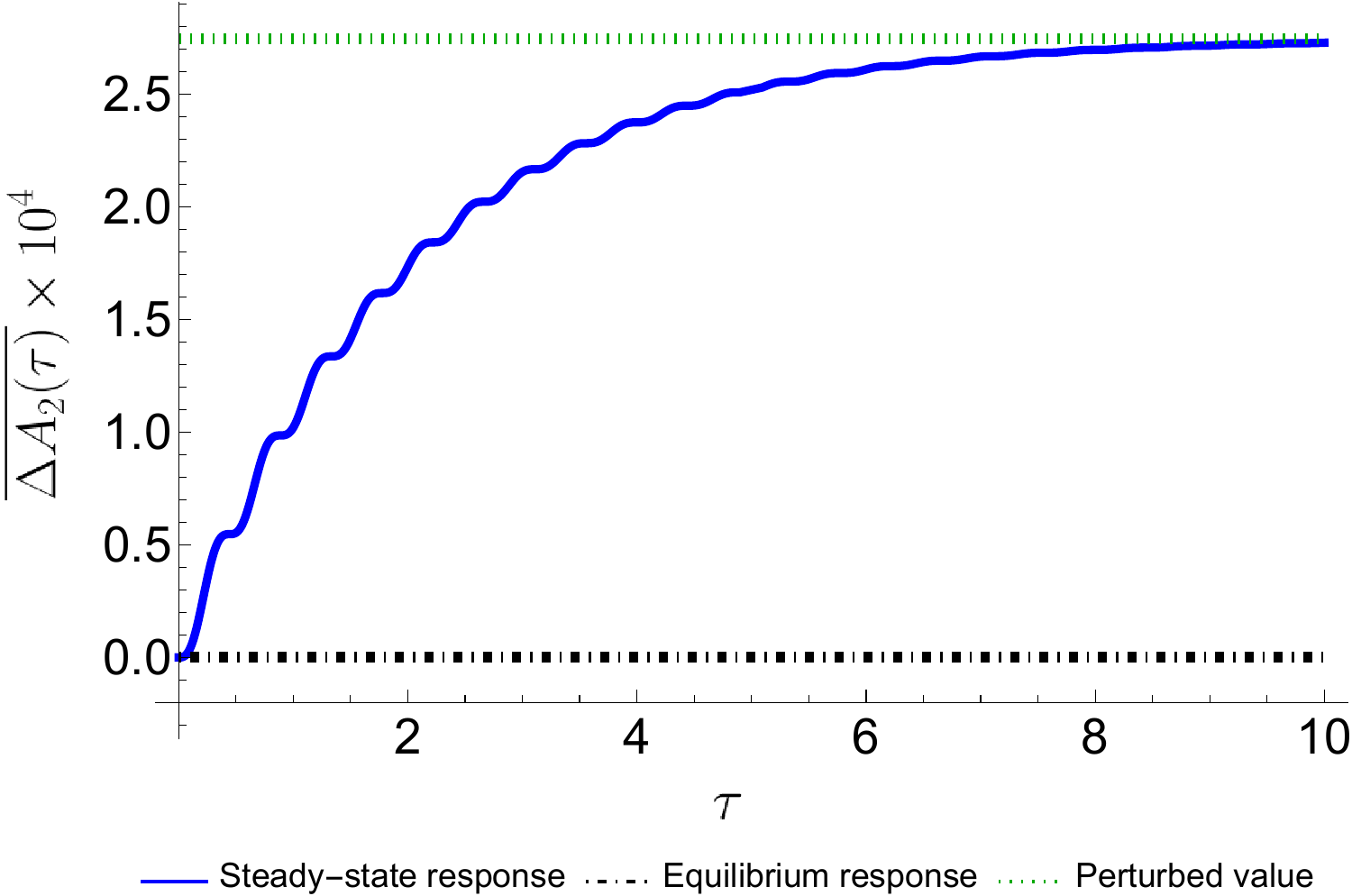}}
\caption{Plot of the response of $A_2=\beta_2\omega_2a_2^\dag a_2$ to a step perturbation to $n_1$. The plots are obtained with the following choices: $\delta=10\omega_1$, $\lambda=5\omega_1$, $\gamma=0.5\omega_1$, $\phi=0.1$, $\beta_1\omega_1=0.1$, and $\beta_2\omega_1=0.001$. The steady-state response approaches the perturbed value, whilst, as expected, in the limit of decoupled oscillators (i.e., $\lambda \to 0$), we obtain a vanishing equilibrium response.}
\label{effectof1on2plot}
\end{figure}

\begin{figure}[t]

\subfloat[\label{subfig:a}]{%
  \includegraphics[width=\linewidth]{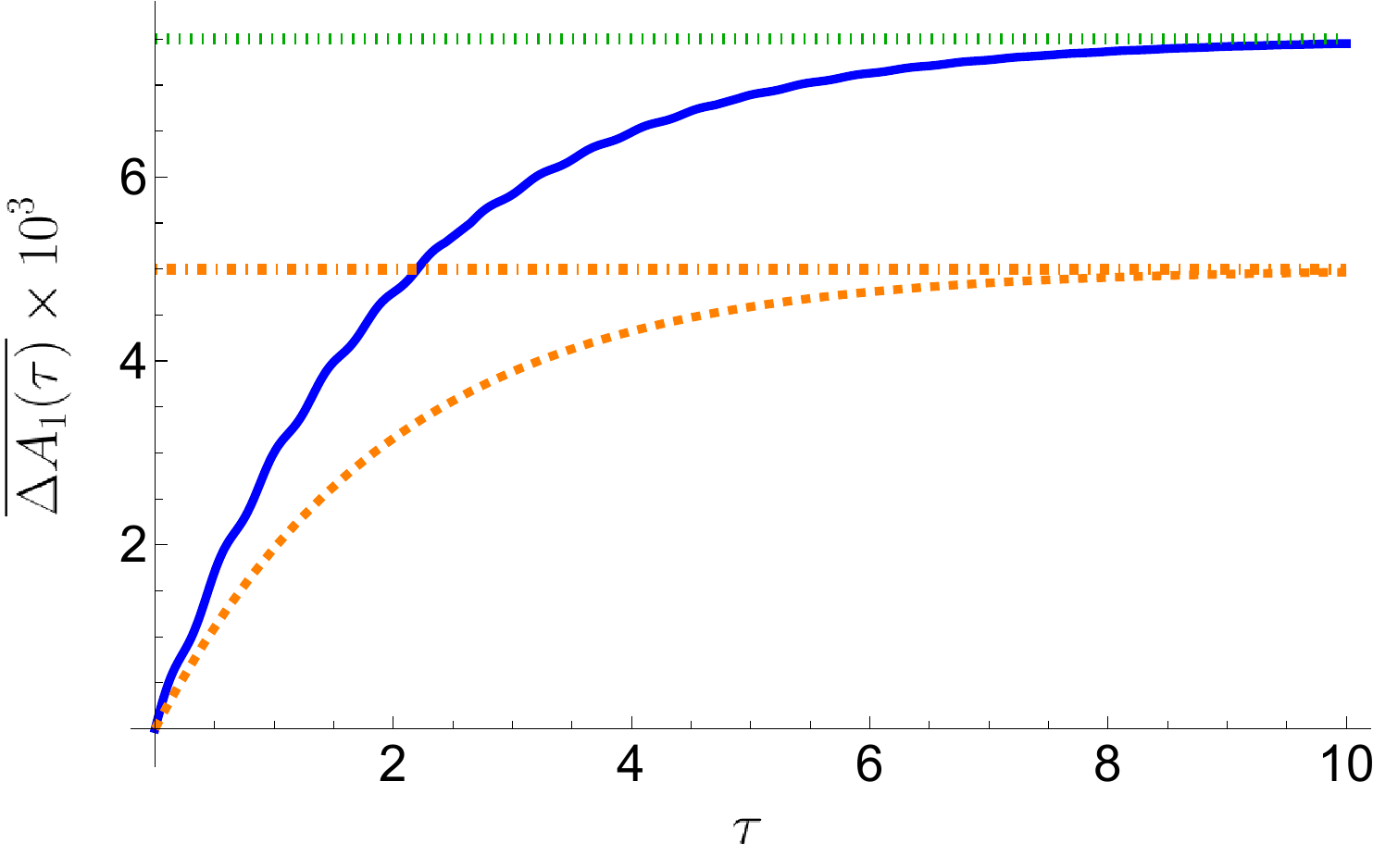}%
}
\vfill
\subfloat[\label{subfig:d}]{%
  \includegraphics[width=\linewidth]{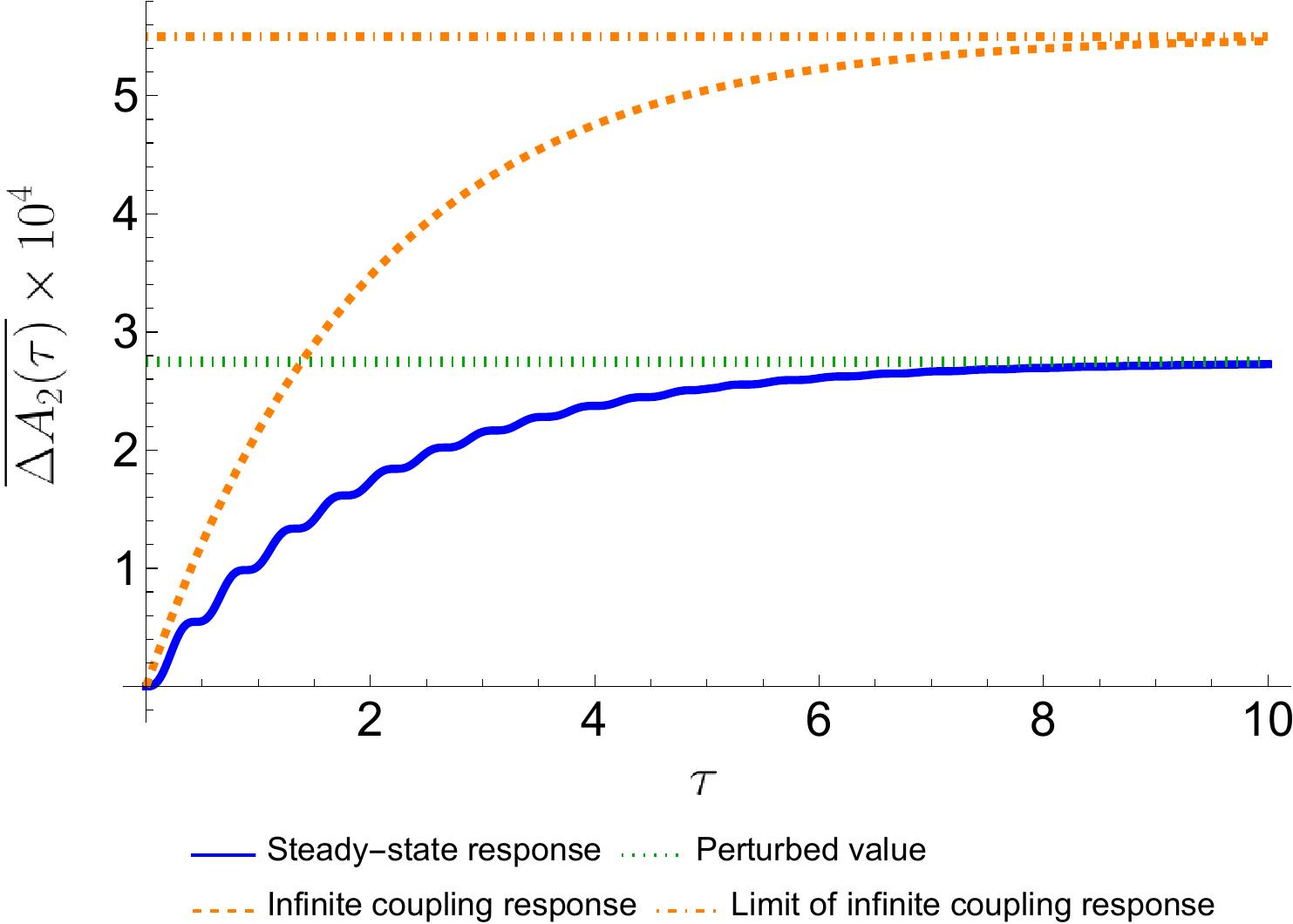}%
}
\caption{Plots of dimensionless energy response to a step perturbation in the number of excitations $n_1$ in the first bath, showing also the response for $\lambda\xrightarrow{}\infty\,$ (infinite coupling response), and the limit of this behaviour as $\tau\xrightarrow{}\infty\,$. The plots are obtained for $\delta=10\omega_1$, $\lambda=5\omega_1$, $\gamma=0.5\omega_1$, $\phi=0.1$, $\beta_1\omega_1=0.1$, and $\beta_2\omega_1=0.001$. Panel (a) shows the response of the first oscillator ($A_1=\beta_1\omega_1a_1^\dag a_1$), while panel (b) refers to the second oscillator ($A_2=\beta_2\omega_2a_2^\dag a_2$).}
\label{fig:comparison}
\end{figure}

The response shown in Fig.~\ref{effectof1on2plot} is qualitatively similar to that in Fig.~\ref{tempplot}, but the magnitude is smaller due to the fact that $\beta_2\omega_2<\beta_1\omega_1$. Furthermore, the perturbation is now mediated by the first harmonic oscillator, which is the one whose bath is perturbed.
As can be deduced from Fig.~\ref{effectof1on2plot}, whenever the coupling between the two oscillators is nonzero, the perturbation applied to the first oscillator affects the second one. As a result, the value of the coupling strength $\lambda$ determines the response of the second oscillator. We provide evidence of this by exploring the limit $\lambda \to 0$, which corresponds to the equilibrium response. As we discussed in Sec.~\ref{sec:SystemThermalBaths}, the equilibrium response of the first oscillator, shown in Fig.~\ref{tempplot}, approaches the maximum value $\overline{\Delta A_1^0(\infty)} = \beta_1\omega_1\phi$, whilst the equilibrium response of second oscillator is zero [cf. Fig.~\ref{effectof1on2plot}]. For a finite value of the coupling constant $\lambda$, i.e., when we look at the steady-state response, the perturbation non-trivially affects the system response.

The difference between the perturbed value for coupled and uncoupled oscillators in Fig.~\ref{tempplot} is related to the perturbed value of the second oscillator in Fig.~\ref{effectof1on2plot} as
\begin{equation}
    \overline{\Delta A_1^0(\infty)}-\overline{\Delta A_1^\lambda(\infty)}=\beta_1\omega_1(\beta_2\omega_2)^{-1}\,\overline{\Delta A_2^\lambda(\infty)}\, ,
\label{eqperturbdiff}
\end{equation}
where $\overline{\Delta A_j^\lambda(\infty)}$ is the perturbed value of the energy of the $j^{\textrm{th}}$ oscillator $(j=1,2)$ for a coupling strength $\lambda$. 
As $\overline{\Delta A_1^0(\infty)}=\beta_1\omega_1\phi$, we find
\begin{equation}
  \phi=\sum_{j=1,2}(\beta_j\omega_j)^{-1}\,\overline{\Delta A_j^\lambda(\infty)}  \, .
\end{equation}
This expression allows us to calculate the perturbed value of one of the oscillators if we know that of the other. This means that we can use the second oscillator as a probe to gain information about the response to a perturbation to the bath of the first. The degree to which the second oscillator mimics the first is determined by the coupling strength $\lambda$. We can add an additional line to the plots in Fig.~\ref{tempplot} and Fig.~\ref{effectof1on2plot} to show the behaviour as $\lambda\xrightarrow{}\infty\,$, illustrated in Fig.~\ref{fig:comparison}. Comparing Fig.~\ref{fig:comparison}~(a) and (b), we see that when $\lambda\xrightarrow{}\infty\,$, the energy response of the first oscillator approaches a value of $\beta_1\omega_1\phi/2$, while -- in the same limit -- the response of the second oscillator tends towards $\beta_2\omega_2\phi/2$. This shows that in the limit of infinite coupling strength, the energy of both oscillators increases by a factor proportional to $\phi/2$. This can be interpreted as an instance of equipartition of energy between the two parties of the composite system.

\section{System interacting with a single squeezed thermal bath}\label{sec:sqbath}

\begin{figure}[t]
    \centering
\scalebox{0.5}{
\includegraphics[]{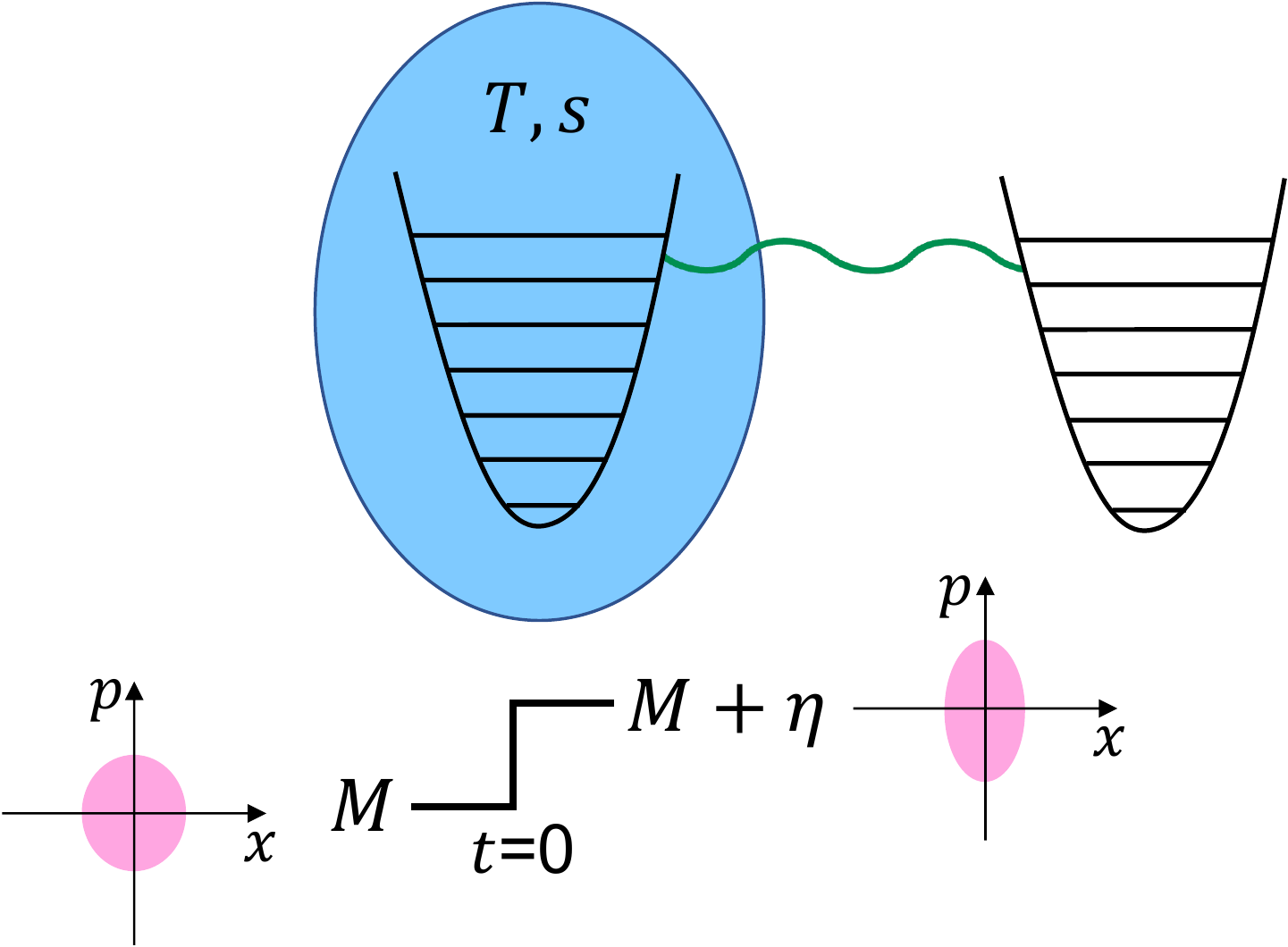}
}
    \caption{Sketch of the second scenario: two coupled quantum harmonic oscillators, one of which is interacting with a squeezed thermal bath characterized by temperature $T$ and squeezing $s$. We investigate the response of the system directly interacting with the bath through the second system, which serves as a probe, when applying a sudden quench to parameter $M$, i.e., $M \to M + \eta$.}
    \label{fig:sqbathdiagram}
\end{figure}

We now investigate a system comprised of two coupled quantum harmonic oscillators, one of which is connected to a squeezed thermal bath, as shown in Fig.~\ref{fig:sqbathdiagram}. The environment is characterized by two parameters, i.e., its temperature $T$ and the squeezing parameter $s$, which can be written in polar form as $s=re^{i\theta}$. The unitary dynamics of this system are described by the Hamiltonian $H_0$ given by Eq.~(\ref{h0}).
In this scenario, the unperturbed dynamics are governed by the following master equation~\cite{WallsMilburn}
\be
\dot{\rho}={\cal L}_0\rho=-i[H_0,\rho]+ \sum_{i=1,2}{\mathcal{D}_i}[\rho] \,,
\label{eq:unpertme2}
\ee
where the dissipator -- acting only on the first system -- comprises the superoperators
\begin{align}
\mathcal{D}_1[\rho]&=
\gamma(N +1)\left( a_1 \rho a_1^\dag -\frac{1}{2}\{a_1^\dag a_1,\: \rho\}\right) \nonumber\\
&+\gamma \:N \:\left( a_1^\dag \rho a_1 -\frac{1}{2}\{a_1 a_1^\dag,\: \rho\}\right), \\
\mathcal{D}_2[\rho]&= \gamma M \left(a_1^\dag \rho a_1^\dag -\frac{1}{2}\{{a_1^\dag}^2, \rho\}\right) +{\rm h.c.} \, ,
\label{eq:MEsqbath}
\end{align}
with
$N=n\left(\cosh^2{r}+\sinh^2{r}\right)+\sinh^2{r}$, 
$M=- \cosh{r}\sinh{r}e^{i\theta}\left(2n+1\right)$, and $n=\left[\exp(\beta\omega_1)-1\right]^{-1}$ is the average number excitations in the bath at inverse temperature $\beta=T^{-1}$. The parameters $M$ and $N$ are not mutually independent as the condition $|M|^2\leq N(N+1)$ must be enforced in order to ensure positivity of the density matrix~\cite{breuer}. 

As in the case discussed in Sec.~\ref{sec:SystemThermalBaths}, we need to find the steady-state $\rho_0$ or, equivalently, the corresponding CM. To do so, we follow a slightly different approach compared to that discussed in Sec.~\ref{sec:SystemThermalBaths}. The core idea is to remap -- resorting to a standard set of correspondence rules~\cite{Gardiner} -- the master equation from Eq.~(\ref{eq:unpertme2}) into a Fokker-Planck equation for the characteristic function of our two-mode system, defined as
\be
\label{eq:chi_def}
\chi(\alpha_1, \alpha_2, t) \equiv \Tr \left \{ D_1(\alpha_1) \otimes D_2(\alpha_2) \rho \right \}\, ,
\ee
where $D_i(\alpha_i) \equiv \exp ( \alpha_i a_i^\dagger - \alpha^*_i a_i)$ is the displacement operator associated to the $i$-th mode \cite{WallsMilburn, Cahill:1969}. The steady-state characteristic function $\chi_0(\alpha_1, \alpha_2)$ is then found by imposing the condition $\dot{\chi}=0$.
By inspection, from $\chi_0(\alpha_1, \alpha_2)$ one can construct
the steady-state CM, i.e., $\mathbf{\Sigma}_0$, written in terms of the annihilation and creation operators. In general, the entries of $\mathbf{\Sigma}$ are defined as
\be
\Sigma_{ij}=\frac{1}{2}\langle\{X_i, X_j^\dag\}\rangle - \langle X_i \rangle \langle X_j^\dag \rangle \,,
\ee
with $\mathbf{X}=(a_1,a_1^\dag,a_2,a_2^\dag)$. Note that $\boldsymbol{\Sigma}$ and $\boldsymbol{\sigma}$ are related by the transformation 
$\boldsymbol{\sigma}=\mathbf{\Lambda\Sigma\Lambda^\dag}$, where $\mathbf{\Lambda}$ is defined as $\boldsymbol{\Lambda} \equiv \bigoplus_{i=1,2} \Lambda_i$ with
\begin{equation}
\Lambda_i = \frac{1}{\sqrt{2}}\begin{pmatrix} &1 & 1 \\ & -i & i \end{pmatrix}.
\label{eq:Lambda}
\end{equation}
The details of the derivation of the steady-state CM are given in the Appendix~\ref{app:ss2}, whereas here we state the final result, i.e.,
\begin{equation}
\mathbf{\Sigma}_0 = 
\begin{pmatrix}
N+\frac{1}{2}
 &D
 &0
 &E^*\\
 D^*
 &N+\frac{1}{2}
 &E
 & 0\\
 0
 &E^*
 &N+\frac{1}{2}
 &F \\
 E
 &0
 &F^*
 &N+\frac{1}{2}
 
    \end{pmatrix}\,,
\label{eq:covmatsqbath_maintxt}
\end{equation}
where the explicit form of the coefficients $D, E$, and $F$ is provided in Appendix~\ref{app:ss2}. These coefficients ultimately depend on the physical parameters characterising the dynamics of the system. In particular, in the limit $\lambda \to 0$, both $E$ and $F$ vanish, as one can immediately check through Eqs.~(\ref{eq:CM_E}) and (\ref{eq:CM_F}).  In other terms, when the two oscillators are uncoupled, the steady-state is given by a block-diagonal CM in the form
\be
\boldsymbol{\Sigma}_0^{\rm eq} = \boldsymbol{\Sigma}^{\rm ND} \oplus  \boldsymbol{\Sigma}^{\rm D},
\ee
where the steady-state CM associated with the first mode, i.e.,
\be 
\boldsymbol{\Sigma}^{\rm ND} = \begin{pmatrix}
N+\frac{1}{2}
 &D
 \\
 D^*
 &N+\frac{1}{2}
 \end{pmatrix} \,,
\ee
is non-diagonal due to the squeezing of the associated bath, whereas the second mode converges to the diagonal CM $\boldsymbol{\Sigma}^{\rm D} = \operatorname{diag} \left ( N + 1/2, N + 1/2 \right )$. Furthermore, in the limit $M \to 0$, we have that the coefficient D vanishes and the system relaxes towards a global thermal state.

\begin{figure}[t]
\center{\includegraphics[width=\linewidth]{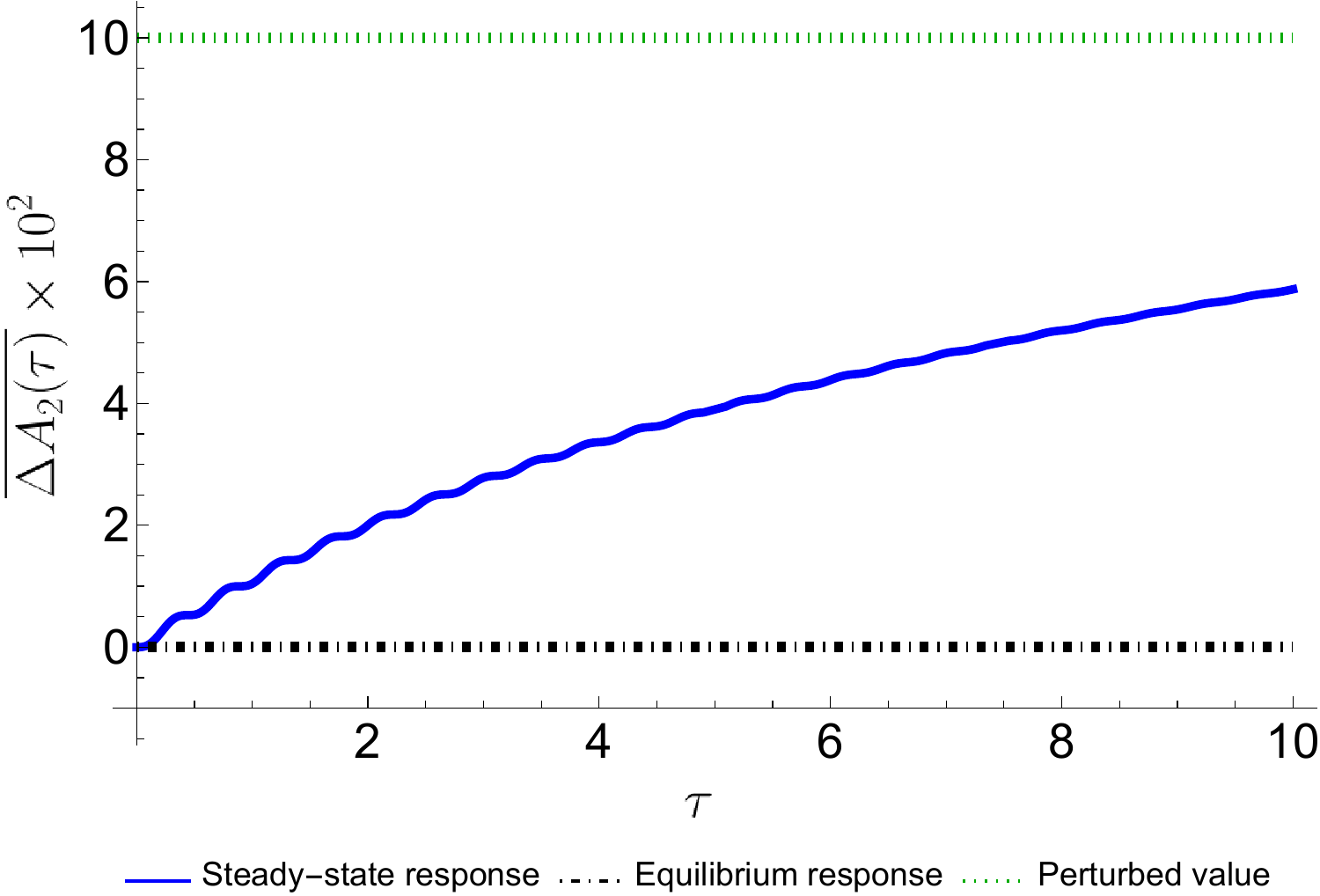}}
\caption{Plot of the dimensionless energy response of the second oscillator to a step perturbation in the squeezing ($M$) of the bath attached to the first oscillator. Curves are obtained for the following choice of the physical parameters: $\delta=10\omega_1$, $\lambda=5\omega_1$, $\gamma=0.5\omega_1$, $\phi=0.1$, $\beta\omega_1=0.1$. The steady-state response converges to the perturbed value. The equilibrium response shows that the energy of the second oscillator is unchanged when the oscillators are not coupled (i.e., when $\lambda \to 0$).}
\label{fig:sqplot}
\end{figure}

\subsection{Perturbation to bath squeezing}
\label{sec:sqpert}

Given this system, we apply the linear response formalism outlined in Sec.~\ref{sec:linresptheory} to study the effect of a perturbation to the squeezing of the first bath. More concretely, the perturbation is described by the superoperator 
\begin{equation} 
\mathcal{L}_1^\eta \rho = \gamma \: \eta (t) \left(a_1^\dag \rho a_1^\dag -\frac{1}{2} \Big\{{
a_1^\dag}^2,\rho\Big\}\right) +\rm{h.c.}\,.
\label{eq:L1_sq}
\end{equation}
We restrict our attention to the case of a sudden perturbation, which we model through the step-like function $\eta(t)= \eta \,\theta(t)$ ($\eta\in\mathbb{C}$). The link between $M$ and $N$ entails that a step-like perturbation on the former will in turn cause a similar perturbation on the latter of the same form as $\mathcal{L}_1^{\phi}$ from Eq.~(\ref{eqL1}). The values of $\eta$ and $\phi$ are related as
\be
\eta = -\frac{2e^{i\theta}\phi\cosh{r}\sinh{r}}{\cosh^2{r}+\sinh^2{r}}\,,
\label{eq:etaphi}
\ee
under the assumption that the squeezing parameter $s=re^{i\theta}$ is not directly affected by the perturbation. The linear response is obtained by accounting for both the perturbations in Eq.~(\ref{eqL1}) and Eq.~(\ref{eq:L1_sq}), i.e.
\begin{equation}
      \overline{\Delta {A_2}(\tau)} = \int_{0}^{\tau} dt\, \text{Tr}\left[\:{A_2}(\tau -t)\left(\mathcal{L}_1^\phi+\mathcal{L}_1^\eta \right)\rho_0\:\right] \,.
\label{eq:sqresp}
\end{equation}
As we aim to use the second oscillator as a probe, we choose the observable ${A_2}(t)=(a_2^\dag a_2)(t)$ to determine the linear response in dimensionless units. The calculation of the time-evolution of ${A_2}(t)$ for this situation follows lines analogous to those previously illustrated,  leading to 
\begin{align}
\label{eq:sq_heis}
(a_2^\dag a_2)(t) = & \tilde{f}(t) a_1^\dag a_1 + [\tilde{g}(t) a_1^\dag a_2 +{\rm h.c.}]+ \tilde{j}(t) a_2^\dag a_2 + \tilde{l}(t),
\end{align}
where $\tilde{f}(t),\tilde{g}(t),\tilde{j}(t)$, and $\tilde{l}(t)$ are defined in Appendix~\ref{app:TimeEvolution}. 

Plugging Eq.~(\ref{eq:sq_heis}) into Eq.~(\ref{eq:sqresp}), we find that the contribution coming from $\mathcal{L}_1^\eta$ is zero, hence the linear response reduces to

\begin{equation}
      \overline{\Delta {A_2}(\tau)} = \int_{0}^{\tau} dt\, \text{Tr}\left[\:{A_2}(\tau -t)\mathcal{L}_1^\phi\rho_0\:\right] \,,
\label{eq:sqrespsimple}
\end{equation}
which is of the same form as that used in Sec.~\ref{sec:SystemThermalBaths}, provided that the time-evolution of ${A_2}(\tau-t)$ is obtained from Eq.~(\ref{eq:sq_heis}), and the steady-state $\rho_0$ is deduced from the CM in Eq.~(\ref{eq:covmatsqbath_maintxt}).
The linear response is finally given by
\begin{equation}
      \overline{\Delta {A_2}(\tau)} = \; \gamma \int_{0}^{\tau} dt \:\phi(t) \tilde{f}(\tau-t)\, ,
\label{eqtempresponsesq}
\end{equation}
which is plotted in Fig.~\ref{fig:sqplot}. 

The perturbed value of the energy of the second oscillator is given by $\phi=\Delta N$, where the change in $N$ is caused by the perturbation of the parameter $M$. At long times, the measurement of the change in the energy of the second oscillator gives a value for $\phi$. Hence we can deduce the change in squeezing, $\eta=\Delta M$, of the bath attached to the first oscillator from Eq.~(\ref{eq:etaphi}). This shows that we can use the second oscillator as a probe of the perturbation to the squeezing of the bath interacting with the first oscillator.

\section{Conclusions and outlook}
\label{sec:conclusions}
We have used the formalism of linear response theory to investigate both unitary and non-unitary perturbations around non-equilibrium steady-states of an open quantum system consisting of two coupled harmonic oscillators. We have calculated the response to simultaneous perturbations to the coupling strength between the two oscillators (i.e., a unitary perturbation), and to the temperature of the thermal bath (i.e., a non-unitary perturbation). We also looked at the effect solely determined by the temperature perturbation, investigating the response of both the oscillators to this. We found that such an approach is effective in  
probing perturbations to the dynamics of the system, including in situations where the latter is affected with a non-equilibrium bath. 

The approach presented in this paper shows how to extend the theory of quantum linear response to non-unitary perturbations affecting the dynamics of open systems. Our study calls for a mathematically consistent formalism able to microscopically account for such perturbations. This would allow quantitative analysis of more general scenarios where the dynamical equations go beyond the usual Markovian approximation, encompassing, e.g., strong-coupling and memory effects.

\acknowledgements

We acknowledge support from the European Union's Horizon 2020 FET-Open project  TEQ (766900), the Horizon Europe EIC Pathfinder project QuCoM (Grant Agreement No.\,101046973), the Leverhulme Trust Research Project Grant UltraQuTe (grant RGP-2018-266), Deutsche Forschungsgemeinschaft (DFG, German Research Foundation) project number BR 5221/4-1, the Royal Society Wolfson Fellowship (RSWF/R3/183013), the UK EPSRC (EP/T028424/1), and the Department for the Economy Northern Ireland under the US-Ireland R\&D Partnership Programme.

\appendix

\section{Steady-state covariance matrix for the first case}
\label{app:ss1}
For the system described in Sec.\ref{sec:SystemThermalBaths}, we find the steady-state using the method set out in Appendix C of Ref.~\cite{MehboudiAcin} to transform some classes of Lindbladian master equations into dynamical equations for the first and second moments. Let us consider a Lindbladian master equation in the form
\begin{equation}
    \dot{\rho}=-i\left[H_0,\rho\right]+\sum_k\Big(L_k\rho L_k^\dag-\frac{1}{2}\big\{L_k^\dag L_k , \rho\big\}\Big) \,,
\label{eq:generalform}
\end{equation}
where the Hamiltonian $H_0$ is quadratic in the quadrature operators, i.e., it can be written in the following matrix form
\be
\label{eq:h0_quadratic}
H_0=\frac{1}{2}\mathbf{Y} \mathbf{G} \mathbf{Y}^T \,,
\ee
where $\mathbf{Y}$ is the vector of quadratures, while the Lindbladian operators are linear, i.e., $L_k=\mathbf{c}_k^T \mathbf{Y}^T$.
It is immediate to see that the master equation given by Eq.~(\ref{eq:unpertme1}) fulfils these requirements, provided that we rewrite the relevant quantities in terms of the quadratures $x_i$ and $p_i$, which are related to the set of creation and annihilation operators, $a_i^\dagger$ and $a_i$, by a linear transformation, as outlined in Sec.~\ref{sec:SystemThermalBaths}. Therefore, the free Hamiltonian of Eq.~(\ref{h0}) reads
\begin{align}
H_0=\frac{\omega_1}{2}(x_1^2+p_1^2)  +\frac{\omega_1+\delta}{2}(x_2^2 &+p_2^2)
+\lambda(x_1x_2+p_1p_2)\,.
\end{align}
The latter can be brought in the form of Eq.~(\ref{eq:h0_quadratic}) by taking $\mathbf{Y}\equiv (x_1,p_1,x_2,p_2)$ as vector of the quadratures, and
\be
\mathbf{G}=\begin{pmatrix} 
    \omega & 0 & \lambda & 0 \\ 
    0 & \omega & 0 & \lambda \\
    \lambda & 0 & \omega+\delta & 0 \\
    0 & \lambda & 0 & \omega+\delta
    \end{pmatrix}\,.
\label{eq:G}
\ee
Comparing the unperturbed master equation in Eq.~(\ref{eqmaster}) with the general form in terms of the jump operators, i.e., Eq.~(\ref{eq:generalform}), we can identify the jump operators as
\begin{equation}
\begin{split}
L_1&=a_1\sqrt{\gamma(n_1+1)}\,;\;\;\; L_2=a_1^\dag \sqrt{\gamma n_1}\,;\;\\
L_3&=a_2\sqrt{\gamma(n_2+1)}\,;\;\;\; L_4= a_2^\dag \sqrt{\gamma n_2}\,.
\end{split}
\end{equation}
Following Ref.~\cite{MehboudiAcin}, we seek the vectors $\mathbf{c}_k$ such that $L_k=\mathbf{c}_k^T \mathbf{Y}^T$; they are given by
\be
\begin{split}
&\mathbf{c}_1^T=\sqrt{\frac{\gamma(n_1+1)}{2}}(1,i,0,0)\,;\;\;\;\; \mathbf{c}_2^T=\sqrt{\frac{\gamma n_1}{2}}(1,-i,0,0)\,;  \\
&\mathbf{c}_3^T=\sqrt{\frac{\gamma(n_2+1)}{2}}(0,0,1,i)\,;\;\;\;\; \mathbf{c}_4^T=\sqrt{\frac{\gamma n_2}{2}}(0,0,1,-i)\,.
\end{split}
\ee
The vectors $\mathbf{c}_k$ allow us to construct the matrix
\begin{multline}
\mathbf{CC}^\dag=\sum_k \mathbf{c}_k\mathbf{c}_k^\dag\\
=\begin{pmatrix}
\frac{\gamma(2n_1+1)}{2} & -\frac{i\gamma}{2} &0 &0 \\ & \\
\frac{i\gamma}{2} & \frac{\gamma(2n_1+1)}{2} &0 &0 \\& \\
0 &0 & \frac{\gamma(2n_2+1)}{2} & -\frac{i\gamma}{2} \\&\\
0&0& \frac{i\gamma}{2} & \frac{\gamma(2n_2+1)}{2}
\end{pmatrix}\,.
\label{eq:CCdag}
\end{multline}
The drift matrix ${\bm\alpha}$ is obtained via \cite{MehboudiAcin}
\be
{\bm\alpha}=-i\mathbf{\Omega}\left(\mathbf{G}-\text{Im}(\mathbf{CC}^\dag)\right)\,,
\label{eq:A}
\ee
where $\text{Im}(\mathbf{M})$ denotes the imaginary part of a matrix $\mathbf{M}$, and $\mathbf{\Omega}$ is the symplectic matrix given by $\mathbf{\Omega} = \mathbf{\Omega}_1 \oplus \mathbf{\Omega}_2$, where
\be \mathbf{\Omega}_j=i
\begin{pmatrix}
0&1\\
-1&0
\end{pmatrix}, \quad j=1,2 \, .
\ee
Substituting Eqs.~(\ref{eq:G}) and (\ref{eq:CCdag}) into Eq.~(\ref{eq:A}) yields
\be
{\bm\alpha}=\begin{pmatrix}
-\frac{\gamma}{2}&\omega&0&\lambda\\
-\omega&-\frac{\gamma}{2}&-\lambda&0\\
0&\lambda&-\frac{\gamma}{2}&\omega+\delta\\
-\lambda&0&-(\omega+\delta)&-\frac{\gamma}{2}
\end{pmatrix}\,.
\ee
The matrix $\mathbf{D}$ is given by~\cite{MehboudiAcin}
\be
\mathbf{D}=\mathbf{\Omega} \,\text{Re}(\mathbf{CC}^\dag)\,\mathbf{\Omega}\,,
\label{eq:D}
\ee
where Re($\mathbf{M}$) denotes the real part of a matrix $\mathbf{M}$. Substituting Eq.~(\ref{eq:CCdag}) into (\ref{eq:D}), the matrix $\mathbf{D}$ turns out to be
\be
\mathbf{D}=\frac{\gamma}{2}
\begin{pmatrix}
2n_1+1 &0&0&0\\
0& 2n_1+1 &0&0\\
0&0& 2n_2+1 &0 \\
0&0&0& 2n_2+1
\end{pmatrix}\,.
\ee

Once the matrices ${\bm\alpha}$ and $\mathbf{D}$ have been obtained, we can determine the steady-state CM, $\boldsymbol{\sigma}_0$, by solving ${\bm\alpha} \boldsymbol{\sigma}+\boldsymbol{\sigma} {\bm\alpha}^T +\mathbf{D}=0$, which gives
\begin{equation}
\boldsymbol{\sigma}_0 = \scalebox{0.9}{
$\zeta\begin{pmatrix}
B+n_1+\frac{1}{2}
 &0
 &-\delta C
 &-\gamma C\\
 0
 &B+n_1+\frac{1}{2}
 &\gamma C
 & -\delta C\\
 -\delta C
 &\gamma C
 &B+n_2+\frac{1}{2}
 &0 \\
 -\gamma C
 &-\delta C
 &0
 &B+n_2+\frac{1}{2}
    \end{pmatrix}$ }  ,
\label{eq:covmat1}
\end{equation}
where $\zeta=\dfrac{\gamma^2+\delta^2}{4\lambda^2+\gamma^2+\delta^2}$, 
    $B=\dfrac{2\lambda^2(n_1+n_2+1)}{\gamma^2+\delta^2}$, and $C=\dfrac{\lambda(n_1-n_2)}{\gamma^2+\delta^2}$.
This matrix can be easily brought in the more compact form of Eq.~(\ref{eqcovmatrix_compact}).

\section{Time-evolution of observables}
\label{app:TimeEvolution}

By following the derivation performed in Ref.~\cite{lutz}, in this Appendix, we will explicitly show how to derive the time-evolution of the observable $a_1^\dag a_1$ in the Heisenberg picture, under the dynamics of the first system with local thermal baths. As shown in Ref.~\cite{lutz}, starting from the adjoint master equation in Eq.~(\ref{eqAdot}), we obtain a set of $10$ coupled differential equations, which can be recast in matrix form as
\begin{equation}
\dot{\mathbf{v}}(t) = \mathbf{M} \cdot \mathbf{v}(t)+\mathbf{w} \, ,
\label{eqode}
\end{equation} 
    where 
    \begin{align}
    \mathbf{v}(t) & =\Big(a_1^\dag a_1 (t), \, a_1^2(t), \, {a_1^\dag}^2(t), \, a_2^\dag a_2(t),\, a_2^2(t),\, \nonumber \\
    & {a_2^\dag}^2(t),\,a_1 a_2(t), a_1 a_2^\dag(t),\,a_1^\dag a_2(t),\, a_1^\dag a_2^\dag(t)\Big)^T, \\
    \mathbf{w} & =\left(n_1 \gamma, 0, 0, n_2 \gamma, 0, 0, 0, 0, 0, 0 \right)^T\,,
    \end{align}
and $\mathbf{M}$ is the $10\times10$ matrix 
\begin{widetext}
\begin{align}
 \mathbf{M} =\begin{pmatrix}
 -\gamma &0&0&0&0&0&0& i\lambda & -i\lambda &0 \\
    0 & -2i\omega_1-\gamma & 0&0&0&0 & -2i\lambda &0&0&0 \\ 0&0& 2i\omega_1 -\gamma &0&0&0&0&0&0&2i\lambda \\ 0&0&0& -\gamma &0&0&0&-i\lambda &i\lambda &0 \\ 0&0&0&0& -2i\omega_2-\gamma &0&-2i\lambda&0&0&0 \\ 0&0&0&0&0& 2i\omega_2 -\gamma &0&0&0& 2i\lambda \\ 0& -i\lambda &0&0& -i\lambda &0& -i\omega_{12} -\gamma &0&0&0 \\ i\lambda &0&0&-i\lambda &0&0&0& -i\Delta\omega -\gamma &0&0\\ -i\lambda &0&0 &i\lambda &0&0&0&0& i\Delta\omega - \gamma &0 \\ 0&0& i\lambda &0&0& i\lambda &0&0&0& i\omega_{12}-\gamma
\end{pmatrix} \, ,
\end{align}
\end{widetext}
with $\omega_{12}=\omega_1+\omega_2$ and $\Delta\omega=\omega_1-\omega_2$. For time-independent $\mathbf{w}$, integration yields
\begin{equation}
    \mathbf{v}(t)=e^{t\mathbf{M}}\mathbf{v}(0)-\left(1- e^{t\mathbf{M}}\right)\mathbf{M}^{-1} \mathbf{w}\,.
\label{eqsoln}
\end{equation}
Taking the first entry of the vector $\textbf{v}$ gives the time-evolution of $a_1^\dag a_1$ as
\begin{align}
    a_1^\dag a_1 (t)  &=  f(t)a_1^\dag a_1 
    + j(t)a_2^\dag a_2
    +[p(t)a_1a_2^\dag +{\rm h.c.}]
    +s(t)\, ,
\label{eqa1daga1}
\end{align}
where the functions in Eq.~(\ref{eqa1daga1}) are defined as
\begin{equation}
\begin{aligned}
    f(t)&=e^{-\gamma t} \left(\delta^2+2\lambda^2+2\lambda^2\cos{zt}\right)/z^2,  \\ 
     j(t)&=2\lambda^2e^{-\gamma t}\left(1-\cos{zt}\right)/z^2, \nonumber \\ 
     p(t)&=\lambda e^{-\gamma t}\left(-\delta+iz\sin{zt}+\delta \cos{zt}\right)/z^2
\end{aligned}
\end{equation}
\begin{multline}
    s(t)=\frac{e^{-\gamma t}}{z^2(\gamma^2+z^2)}\Bigg\{\Big[(\gamma^2+z^2)\Big(\delta^2 n_1 +2\lambda^2 (n_1 +n_2)\Big)\\
    -z^2 e^{\gamma t}\Big(n_1(\gamma^2+\delta^2)
    +2\lambda^2(n_1+n_2)\Big)\Big] \\
    +2\gamma\lambda^2 (n_1-n_2)(\gamma \cos{zt}-z\sin{zt})\Bigg\}\, ,
\label{eqfgh}  
\end{multline}
with $z^2=\delta^2+4\lambda^2.$ 

We use a very similar method to find the time-evolution of the observable $a_2^\dag a_2$ for the second system, interacting with a squeezed bath, as detailed in Sec. \ref{sec:sqbath}. We find that
\begin{equation}
(a_2^\dag a_2)(t) = \tilde{f}(t) a_1^\dag a_1 + [\tilde{g}(t) a_1^\dag a_2 +{\rm h.c.}]+ \tilde{j}(t) a_2^\dag a_2 + \tilde{l}(t),
\label{eq:a2daga2}
\end{equation}
where the functions $\tilde{f}(t),\tilde{g}(t),\tilde{j}(t)$, and $\tilde{l}(t)$ are defined as

\begin{widetext}
\begin{equation}
\begin{aligned}
\tilde{f}(t)& {=} \frac{8 \lambda^2 }{\xi} e^{-\frac{\gamma t}{2}} \left[  \cosh \left ( \frac{t}{2} \sqrt{\frac{\zeta^2 {+} \xi}{2}}\right )
 {-} \cosh \left ( \frac{t}{2} \sqrt{\frac{\zeta^2 {-} \xi}{2}}\right ) \right],\\
    \tilde{g}(t)&=\frac{\lambda}{\sqrt{2}\,\xi}\Bigg[-i\frac{e^{-[\frac{\gamma}{2}+\frac14\sqrt{2(\zeta^2-\xi)}]t}}{\sqrt{\zeta^2-\xi}}
    \Bigg(16\lambda^2
    +\xi-(\gamma-2i\delta)\left(\gamma-2i\delta-\sqrt{2(\zeta^2-\xi)}\right)\Bigg)\\ 
    &+\frac{e^{-[\frac{\gamma}{2}-\frac14\sqrt{2(\zeta^2-\xi)}]t}}{\sqrt{\zeta^2-\xi}}\Bigg(16\lambda^2+\xi-(\gamma-2i\delta)\Big(\gamma-2i\delta+\sqrt{2(\zeta^2-\xi}\Big)\Bigg)\\
    &+i\frac{e^{-[\frac{\gamma}{2}-\frac14\sqrt{2(\zeta^2+\xi)}]t}}{\sqrt{\zeta^2+\xi}}
    \Bigg(-16\lambda^2
    +\xi+(\gamma-2i\delta)\left(\gamma-2i\delta+\sqrt{2(\zeta^2+\xi)}\right)\Bigg)\\
    &-i\frac{e^{-[\frac{\gamma}{2}+\frac14\sqrt{2(\zeta^2+\xi)}]t}}{\sqrt{\zeta^2+\xi}}
    \Bigg(-16\lambda^2
    +\xi+(\gamma-2i\delta)\left(\gamma-2i\delta-\sqrt{2(\zeta^2+\xi)}\right)\Bigg)
    \Bigg],\\
    \tilde{j}(t)&=\frac{1}{8\sqrt{2}\xi}\Bigg[\frac{2 e^{-\left(\frac{\gamma}{2} +\frac14\sqrt{2(\zeta ^2-\xi)}\right)t}}{\sqrt{\zeta^2-\xi}}\Bigg(2 \gamma ^3- \gamma ^2 \sqrt{2(\zeta ^2-\xi)}-2 \gamma  \left(32 \lambda ^2+\xi-4 z^2\right)+\sqrt{2(\zeta ^2-\xi)}\left(16 \lambda ^2+\xi-4 z^2\right)\Bigg)\\
    &+\frac{2e^{\left(-\frac{\gamma}{2} +\frac14\sqrt{2(\zeta ^2-\xi)}\right)t}}{\sqrt{\zeta^2-\xi}}\Bigg(2 \gamma ^3- \gamma ^2 \sqrt{2(\zeta ^2-\xi)} +2 \gamma  \left(32 \lambda ^2+\xi-4 z^2\right)+
    \sqrt{2(\zeta ^2-\xi)}\left(16 \lambda ^2+\xi-4 z^2\right)\Bigg)\\
    &+\frac{e^{-\left(\frac{\gamma}{2} +\frac14\sqrt{2(\zeta ^2+\xi)}\right)t}}{\sqrt{\zeta^2+\xi}}\Bigg(64 \gamma  \lambda ^2+2
    \left(\sqrt{2(\zeta ^2+\xi)}-2 \gamma \right) \left(\gamma ^2-16 \lambda ^2+\xi+4 z^2\right)\Bigg)\\
    &+\frac{e^{\left(-\frac{\gamma}{2} +\frac14\sqrt{2(\zeta ^2+\xi)}\right)t}}{\sqrt{\zeta^2+\xi}}\Bigg(-64 \gamma  \lambda ^2+2  
    \left(\sqrt{2(\zeta ^2+\xi)}+2 \gamma \right) \left(\gamma ^2-16 \lambda ^2+\xi+4 z^2\right)\Bigg)
    \Bigg],\\
   \tilde{l}(t)&=n+\frac{1}{4\gamma\xi^2}n e^{-\frac{1}{2}\gamma t} \Bigg[\sqrt{2} \Bigg(\sqrt{\zeta ^2-\xi} \sinh \left(\frac{t \sqrt{\zeta ^2-\xi}}{2 \sqrt{2}}\right)\Big(\gamma ^2\left(-\gamma ^2+64 \lambda ^2+\xi\right)-16 z^4+4 z^2 \left(\xi-2 \gamma ^2\right)\Big)\\
   &-\sqrt{\zeta ^2+\xi}\sinh \left(\frac{t \sqrt{\zeta ^2+\xi}}{2 \sqrt{2}}\right) \Big(\gamma ^2\left(\gamma ^2-64 \lambda ^2+\xi\right)+16 z^4+4 z^2 (2 \gamma ^2 +\xi)\Big)\Bigg)\\
   &-2 \gamma  \cosh \left(\frac{t \sqrt{\zeta ^2-\xi}}{2 \sqrt{2}}\right) \Big(\gamma ^2 \left(\gamma ^2-64 \lambda ^2-\xi\right)+16 z^4-4 z^2 (\xi-2 \gamma ^2)\Big)\\
   &-2 \gamma  \cosh \left(\frac{t \sqrt{\zeta ^2+\xi}}{2 \sqrt{2}}\right) \Big(\gamma ^2 (\gamma ^2-64 \lambda ^2+\xi)+16 z^4+4 z^2 \left(2 \gamma ^2+\xi\right)\Big)\Bigg] \, ,
\end{aligned}
\end{equation}
\end{widetext}
with $\zeta^2 = \gamma^2 - 4 z^2$, $\xi^2 = (4z^2 + \gamma^2)^2 - 64 \gamma^2 \lambda^2$ and $z^2=\delta^2+4\lambda^2$.


\section{Steady-state covariance matrix for the second case}
\label{app:ss2}

The dynamical system discussed in Sec.~\ref{sec:sqbath} is governed by a master equation that does not resemble the general form given in Eq.~(\ref{eq:generalform}). Therefore, we use a different method for finding the steady-state CM. As described in Ref.~\cite{WallsMilburn}, the master equation can be recast in the form of a Fokker-Planck (FP) equation for the two-mode characteristic function $\chi(\alpha_1,\alpha_2, t)$ [cf.  Eq.~(\ref{eq:chi_def})] through the use of standard quantum optics correspondence relations. This allows us to write Eq.~(\ref{eq:unpertme2}) as
\begin{equation}
\label{eq:FP}
\dot{\chi}(\alpha_1,\alpha_2, t)= \left(\,\sum_{k=U,D_1,D_2}\mathcal{F}_{k}\right)\chi(\alpha_1,\alpha_2, t),
\end{equation}
with
\begin{equation}
    \begin{aligned}
&\mathcal{F}_U = i \left (\alpha_1 \omega_1 + \lambda \alpha_2 \right ) \partial_{\alpha_1} + i \left [ (\omega_1 + \delta) \alpha_2 + \lambda \alpha_1 \right] \partial_{\alpha_2} + \rm{h.c.},\\ 
&\mathcal{F}_{D_1} =  - \frac{\gamma}{2} \left[ (2 N +1 ) |\alpha_1|^2  + \alpha_1\partial_{\alpha_1} + \alpha_1^*\partial_{\alpha_1^*}  \right],\\ 
&\mathcal{F}_{D_2} = -\left(M^*\alpha_1^2+M{\alpha_1^*}^2
\right).
\label{eq:chidot}
\end{aligned}
\end{equation}
We consider the following ansatz for the form of the steady-state characteristic function  
\be
\chi_0(\alpha_1,\alpha_2)=\frac{1}{\pi\sqrt{\det\mathbf{\Sigma}_0}}\exp\left(i\mathbf{u
}^\dag\boldsymbol{\bar{X}}_0^T-\frac{1}{2}\mathbf{u}^\dag\mathbf{\Sigma}_0\mathbf{u}\right),
\label{eq:ansatz}
\ee
where we have introduced $\mathbf{u}=(\alpha_1,\alpha_1^*,\alpha_2,\alpha_2^*)^T$, the vector $\boldsymbol{\bar{X}}_{0} = \langle \boldsymbol{X} \rangle_{0}$ of first momenta of the steady-state, and the steady-state covariance matrix $\mathbf{\Sigma}_0$. By solving the algebraic equations stemming from enforcing the steady-state condition $\dot{\chi}(\alpha_1,\alpha_2,t)=0$, we find
\begin{equation}
\mathbf{\Sigma}_0 = 
\begin{pmatrix}
N+\frac{1}{2}
 &D
 &0
 &E^*\\
 D^*
 &N+\frac{1}{2}
 &E
 & 0\\
 0
 &E^*
 &N+\frac{1}{2}
 &F \\
 E
 &0
 &F^*
 &N+\frac{1}{2}
 \end{pmatrix}\,,
\label{eq:covmatsqbath_app}
\end{equation}
where
\be
\label{eq:CM_D}
D=\frac{M\gamma\left\{-2i\lambda^2+(\omega_1+\delta)\left[\gamma+2i\left(2\omega_1+\delta\right)\right]\right\}}{\left[-2i\lambda^2+(\gamma+2i\omega_1)(\omega_1+\delta)\right]\left[\gamma+2i\left(2\omega_1+\delta\right)\right]}\,,
\ee
\be
\label{eq:CM_E}
E=\frac{2iM^*\gamma\lambda\left(\omega_1+\delta\right)}{\left[2i\lambda^2+(\gamma-2i\omega_1)(\omega_1+\delta)\right]\left[\gamma-2i\left(2\omega_1+\delta\right)\right]}\,,
\ee
\be
\label{eq:CM_F}
F=\frac{2iM\gamma\lambda^2}{\left[-2i\lambda^2+(\gamma+2i\omega_1)(\omega_1+\delta)\right]\left[\gamma+2i\left(2\omega_1+\delta\right)\right]}\,.
\ee


\section{Exact dynamic response}
\label{app:exactresp}

In the case of applying one perturbation to temperature or squeezing, the linear response, $\overline{\Delta {A}(\tau)}$, is identical to the exact dynamic response, which we shall denote $\overline{\Delta {A}_E(\tau)}$. We calculate the exact dynamic response in a similar way to the calculation for the perturbed value [cf.  Eq.~(\ref{eq:pertval2}], but this time looking for the change in energy as a function of time rather than at the steady-state. For instance, to find the exact dynamic response of the energy of the first oscillator to a perturbation in temperature, we choose $A=\beta_1\omega_1 a_1^\dag a_1$ and calculate
\be
\overline{\Delta {A_E}(\tau)}=\beta_1\omega_1\left(\left\langle a_1^\dag a_1 (\tau,n_1+\phi)\right\rangle -\left\langle a_1^\dag a_1(\tau, n_1)\right\rangle \right)\,.
\label{eq:exactresp}
\ee
In the expansion of $a_1^\dag a_1(t)$ given in Eq.~(\ref{eqa1daga1}), $s(t)$ is the only term that depends on $n_1$. Hence, Eq.~(\ref{eq:exactresp}) simplifies to
\be
\overline{\Delta {A_E}(\tau)}=\beta_1\omega_1\left[s(\tau,n_1+\phi)-s(\tau,n_1)\right]\,,
\ee
which exactly matches the linear response plotted in Fig.~\ref{tempplot}. 

Similarly, for the perturbation in squeezing, note that $\tilde{l}(t)$ is the only term that depends on the perturbed parameter, i.e., $n$, in the expansion of our observed quantity [cf.  Eq.~(\ref{eq:a2daga2})], which is $A=a_2^\dag a_2$ in this case. Therefore, the exact dynamic response of the second oscillator to a perturbation in the squeezing of the first bath is given by
$\overline{\Delta {A}_E(\tau)}=\tilde{l}(\tau, n+\phi)-\tilde{l}(\tau,n)$.
Again, the linear response perfectly reproduces this exact dynamic response.

\end{document}